\documentstyle[twocolumn,aps,epsf]{revtex}

\newcommand{\boldvec}[1]{\mbox{\boldmath${#1}$}}

\begin{document}

\draft

\wideabs{

\title{Rotational Dynamics of Vortices in Confined Bose-Einstein
Condensates}

\author{S. A. McGee and M. J. Holland}

\address{JILA and Department of Physics, University of Colorado,
  Boulder CO 80309-0440, USA.}

\date{\today}

\maketitle

\begin{abstract}
We derive the frequency of precession and conditions for stability for
a quantized vortex in a single-component and a two-component
Bose-Einstein condensate. The frequency of precession is proportional
to the gradient of the free energy with respect to displacement of the
vortex core. In a two-component system, it is possible to achieve a
local minimum in the free energy at the center of the trap. The
presence of such a minimum implies the existence of a region of
energetic stability where the vortex cannot escape and where one may
be able to generate a persistent current.
\end{abstract}
}

\section{Introduction}
Ever since Bose-Einstein condensation was first observed in a dilute
atomic gas~\cite{Anderson}, considerable attention has been devoted to
understanding its rotational properties. Numerous connections have
already been made between superfluid phenomena in the weakly
interacting atomic gases and superfluid phenomena observed in
condensed matter systems (e.g.~\cite{critical}). Recent demonstrations
of the ability to create and observe singly quantized
vortices~\cite{Williams,Matthews}, and vortex arrays~\cite{ens}, have
opened up the possibility for investigations of vortex
dynamics. Vortex dynamics in this context refers to the motion of
topological defects within the superfluid.

A quantized vortex is represented by a singularity in a superfluid
order parameter. Since the order parameter must be single-valued, the
condition for quantized circulation derives from the fact that the
superfluid phase $\phi$ must undergo a $2\pi l$ change around any
closed contour, where $l$ is an integer. Consequently, the velocity
field in a superfluid is irrotational everywhere except at a vortex
defect or singularity. The density at the defect is zero so that the
current density vanishes even though the superfluid velocity
$\boldvec{v}$ at that coordinate is infinite. The superfluid velocity
is found from the gradient of the phase by
$\boldvec{v}=(\hbar/m)\nabla\phi$, where $m$ is the
mass~\cite{Fetter}. The vortex core is the region around the defect in
which the density falls from its asymptotic value to zero. The spatial
scale for the core is characterized by the ``healing''
length~$\zeta=(8\pi \rho a)^{-1/2}$, where $\rho$ is the local
superfluid number density and $a$ is the $s$-wave scattering length
characterizing the interactions.

It is anticipated that experimental observations on vortex
dynamics~\cite{dynamics,dynamics2} in dilute atomic Bose-Einstein
condensates should agree quantitatively with predictions of the
mean-field theory. External parameters such as the trap properties and
the density and composition of the cloud can typically be controlled
with a high degree of precision. The positions of vortex cores can be
directly observed by imaging techniques rather than indirectly
inferred. These features, along with possibility for multiple
condensate components to be simultaneously present, make the dilute
gas experiments ideal candidates for studies of vortex dynamics.

In this paper, we derive analytic solutions of the equations of motion
for vortices with unit angular momentum and make comparison with full
numerical simulations. The paper is outlined as follows. We begin by
examining in detail a model system composed of a straight line vortex
in a uniform density single-component superfluid which is confined in
an infinite cylindrical vessel. Analytic solutions are presented for
this well-known system for the motion of the vortex defect and
connections are made with numerical calculations of the free
energy. We extend these results to consider the implications of a
harmonic confining potential with the associated quadratic
Thomas-Fermi density envelope.

Analysis of such model systems allows us to elucidate the systematic
method for generalization to more complicated systems in which
analytic solutions are not easily tractable. In particular we
investigate numerically the two-component condensate (relevant to
current experiments at JILA for example) in which the effects of
``buoyancy'' must be considered. Buoyancy in this context is used to
refer descriptively to a net mean-field force on a constituent of a
multi-component condensate due to the various interspecies and
intraspecies interaction parameters.

\section{Mean-Field Theory}

For a single-component condensate, we derive the motion of the vortex
defect by solving the evolution of the superfluid order parameter
$\Phi(\boldvec{r})$ according to the Gross-Pitaevskii equation
\begin{equation}
i\hbar \frac{d\Phi(\boldvec{r})}{dt} = \bigl(
-\frac{\hbar^2}{2m}\nabla^2+ V(\boldvec{r}) + g|\Phi(\boldvec{r})|^2
\bigr) \Phi(\boldvec{r})
\end{equation}
where $V(\boldvec{r})$ is the external potential and
$g=4\pi\hbar^2a/m$. The initial condition is found by evaluating the
lowest energy solution to the time-independent form of the
Gross-Pitaevskii equation consistent with a given total number of
atoms and a given position of the vortex defect. This requires
minimizing the free energy $E$ where
\begin{eqnarray}
E&=&\int d^3\boldvec{r}\,
\Bigl(\frac{\hbar^2}{2m}|\nabla\Phi(\boldvec{r})|^2
+V(\boldvec{r})|\Phi(\boldvec{r})|^2
+\frac{g}{2}|\Phi(\boldvec{r})|^4\Bigr)
\end{eqnarray}
with the imposed constraints.

In the case of a two-component condensate, these equations are
modified to account for the fact that the order parameter is a spinor
$\bigl(\Phi_1(\boldvec{r}),\Phi_2(\boldvec{r})\bigr)$. The free energy
is then given by
\begin{eqnarray}
E&=&\int d^3\boldvec{r}\,
\Bigl(\frac{\hbar^2}{2m}\bigl(|\nabla\Phi_1(\boldvec{r})|^2
+|\nabla\Phi_2(\boldvec{r})|^2\bigr) \nonumber\\ &&+
V_1(\boldvec{r})|\Phi_1(\boldvec{r})|^2 +
V_2(\boldvec{r})|\Phi_2(\boldvec{r})|^2 \nonumber\\ && +
\frac{g_{11}}{2}|\Phi_1(\boldvec{r})|^4
+g_{12}|\Phi_1(\boldvec{r})\Phi_2(\boldvec{r})|^2
+\frac{g_{22}}{2}|\Phi_2(\boldvec{r})|^4\Bigr)
\label{gpeforone}
\end{eqnarray}
In the case of $^{87}$Rb, which we will focus on here, the relevant
matrix elements which characterize the interspecies and intraspecies
interactions of the condensates in the applicable hyperfine states
have similar values as indicated by the relationships
$g_{11}=0.97g_{12}$ and $g_{22}=1.03g_{12}$.

For both the single-component case and the two-component case, we
minimize the free energy using a steepest descents
algorithm~\cite{Williams:thesis}. This involves propagating the
Gross-Pitaevskii equation in imaginary time (by making the simple
substitution $t\rightarrow -it$), and adjusting the order parameter at
each step in the propagation to account for the imposed
constraints. The algorithm is straightforward to implement and
converges efficiently on the self-consistent lowest energy solution.

The constraints we impose depend on the symmetry of the state we wish
to investigate. Since the imaginary time propagation does not preserve
normalization, after each numerical step it is necessary to
renormalize the order parameter to give the correct total number of
atoms~$N$. For a single-component condensate, the condition is $N=\int
d^3\boldvec{r}\,|\Phi(\boldvec{r})|^2$. Placing the vortex defect at a
given location is implemented by imposing a phase pattern about the
chosen point. A single unit of circulation requires a $2\pi$ phase
wrap in the order parameter. We therefore enforce that at each
numerical step the order parameter has a complex phase argument of
$\exp(i\theta)$ about the vortex defect, where $\theta$ is the usual
counterclockwise angle measured in the plane perpendicular to the
vortex line.

For a two-component condensate the method used is similar. Given a
proportion $p$ of atoms in the first condensate component, the
normalization constraints are $Np=\int
d^3\boldvec{r}\,|\Phi_1(\boldvec{r})|^2$ and $N(1-p)=\int
d^3\boldvec{r}\,|\Phi_2(\boldvec{r})|^2$. We adopt the convention that
the component~2 is the state which will contain the single vortex
line, while component~1 will contain no vortices and will therefore
tend to fill the vortex core of component~1 to plug the hole in the
density. In order to approximate this situation using the method of
steepest descents, we enforce that the phase of $\Phi_1(\boldvec{r})$
is spatially uniform at each numerical step. The phase of component~2
is fixed at each numerical step to give a $2\pi$ phase winding around
the chosen position of the defect in the same manner described for a
single-component condensate.

For simplicity, we take the system to be translationally invariant
along the dimension parallel to the vortex line. This allows us to
solve the Gross-Pitaevskii equation in two dimensions rather than in
three dimensions. To allow for comparison with experiment, the number
of atoms per unit length in the two-dimensional Gross-Pitaevskii
equation should take the value which reproduces the same chemical
potential as the equivalent total number of atoms in the real
three-dimensional system.

\section{Single-Component Condensate}

We begin our conceptual treatment of vortex dynamics in model systems
with simple cases involving the motion of a quantized vortex line in a
cylinder.

\subsection{Uniform Density Distribution}

Consider a uniform density superfluid confined in an infinite
cylindrical vessel of radius $R$. A vortex is placed in the superfluid
with the vortex line displaced from the cylinder axis by
$\boldvec{r_0}$. We define the circulation $\boldvec{\kappa}$ in the
usual way to have magnitude $2\pi l\hbar/m$ and to be aligned parallel
to the vortex line with direction determined according to the usual
right-hand rule applied to the superfluid flow. The velocity field can
be found using an image vortex argument~\cite{hess}. The effect of the
boundary conditions at the cylinder walls is to require that the
perpendicular component of the superfluid velocity is zero at the
surface.  This condition is satisfied by considering a formally
equivalent situation of a uniform fluid of infinite extent with an
additional image vortex of opposite circulation placed at
$\boldvec{r_1}=(R/r_0)^2\,\boldvec{r_0}$. The velocity field for this
situation is:
\begin{eqnarray}
\boldvec{v}(\boldvec{r})&=&\frac{\boldvec{\kappa}}{2\pi}\times
\Bigl(\frac{\boldvec{r} - \boldvec{r_0}}
{|\boldvec{r}-\boldvec{r_0}|^2}-\frac{ \boldvec{r}-\boldvec{r_1}}
{|\boldvec{r}-\boldvec{r_1}|^2}\Bigr).
\end{eqnarray}
The motion of the vortex defect at $\boldvec{r_0}$ is found by
computing the superfluid flow at that coordinate, which is due solely
to the image vortex contribution:
\begin{equation}
\boldvec{v}(\boldvec{r_0})=\frac{\boldvec{
\kappa}\times\boldvec{r_0}}{2\pi(R^2-r_0^2)}
\label{vfield}
\end{equation}
The angular frequency of precession of the vortex defect about the
center axis of the cylinder is therefore given by
$\boldvec{\omega}=\boldvec{\kappa}/2\pi(R^2-r_0^2)$. Note that the
angular precession direction is always in the same sense as the
circulation.

An alternative approach (but equivalent method) for finding the rate
and direction of precession of the vortex defect is to compute the
free energy due to the attractive interaction between the real vortex
and the image vortex. Ignoring the kinetic energy within the core of
the vortex line, the free energy per unit length is given
by~\cite{hess}
\begin{equation}
E=\frac{\rho\kappa^2m}{4\pi} \log\Bigl[\frac{R^2-r_0^2}{R\zeta}\Bigr]
\label{free}
\end{equation}
The defect velocity $\boldvec{v}$ is then found from the solution of
\begin{equation} 
\rho m(\boldvec{\kappa}\times\boldvec{v})=\nabla E
\label{findmotion}
\end{equation}
where $\nabla E$ is the gradient of the free energy with respect to
the location of the defect, i.e.\ $\boldvec{r_0}$. The importance of
this alternative approach is that, for a complex system, the free
energy surface $E$ can be calculated numerically. Consequently, the
implication is that one may find the gradient of the free energy with
respect to displacement of the defect as a model to infer the behavior
of the vortex dynamics even when a simple analytic expression such as
Eq.~(\ref{free}) cannot easily be derived.

\begin{figure}
\begin{center}
  \epsfysize=65mm \epsfbox{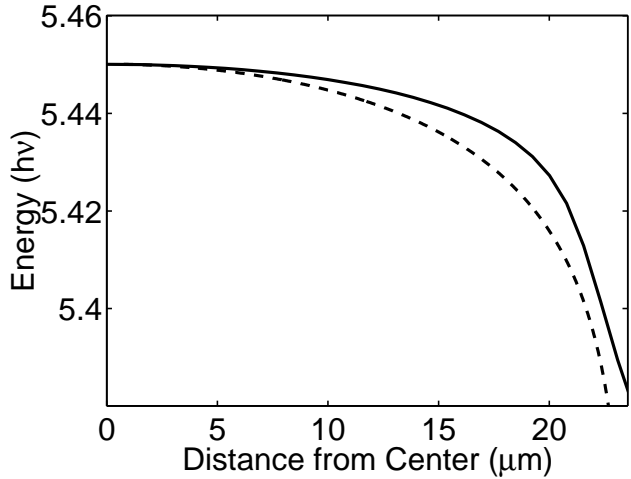}
\end{center}
\caption{Comparison of the approximate analytic expression for the
  free energy given in Eq.~(\ref{free}) (dashed line), which ignores
  the kinetic energy associated with the core, with the exact free
  energy found from the numerical solution to the Gross-Pitaevskii
  equations (solid line) as the amount of displacement $r_0$ is
  varied. Parameters used were $\nu=7.8$~Hz (for convenience for
  comparison with later figures), $N=4\times10^5$, $a=100a_0$ where
  $a_0$ denotes the Bohr radius, $m$ is the mass of $^{87}$Rb, and
  $R=23.5\,\mu m$.}
\label{HomoGPE}
\end{figure}

In order to illustrate this point, in Fig.~\ref{HomoGPE} we show the
results of a numerical solution of the free energy for a displaced
vortex in a cylinder. The external potential $V(\boldvec{r})$ for this
case changes abruptly from zero inside the cylinder to infinity at the
walls. The resulting superfluid density is approximately uniform
except within a small distance of the surfaces as characterized by the
healing length. The exact numerical results are compared with the
approximate analytic free energy expression given in
Eq.~(\ref{free}). According to the analytic expression, the free
energy diverges at the edge of the cloud, since the boundary effects
associated with the core size have not been taken into account. Apart
from edge effects, the analytic expression for the free energy agrees
very well with the total energy found from the numerical solution of
the free energy.
 
At zero temperature there is no energy dissipation and the motion of
the vortex defect is along an equipotential of the free energy, which
is circular in this case. However, in the presence of dissipation, the
defect propagates to regions of lower free energy. The vortex and
image vortex attract each other in the uniform fluid since they have
the opposite sign for the circulation. Consequently, for the situation
considered in Fig.~\ref{HomoGPE} the vortex will spiral out and
eventually annhilate with the image vortex at the surface of the
cylinder. At finite temperature, dissipation is generated by
collisions between superfluid atoms and atoms from the non-condensed
component of the cloud.
 
\subsection{Thomas-Fermi Density Distribution}

We now replace the superfluid of uniform density by the quadratic
Thomas-Fermi density envelope. This is a good approximation to the
density distribution which results from a harmonic confining potential
and allows us to make contact with an experimentally more relevant
situation for dilute atomic Bose-Einstein condensates. For this
distribution, the density falls to zero at the Thomas-Fermi radius $R$
in a smooth and continuous manner so the importance of edge effects is
reduced.

The free energy surface results from the velocity field associated
with the minimum energy configuration at each value of the radial core
displacement and this may be calculated analytically in a hydrodynamic
approach~\cite{Fetter,Lundh}. This approach is in the same spirit as
the method used for the derivation of the free energy in
Eq.~(\ref{free}) in that the kinetic energy associated with the core
region, and the spatial dependence of the trapping potential and
mean-field potential across the vortex core are neglected. Taking into
account the Thomas-Fermi density distribution in this way gives the
free energy expression
\begin{eqnarray}
E&=&\frac{\rho_0\kappa^2m}{8\pi}\biggl[ \frac{ R^2-r_0^2}{R^2}
\ln\frac{R^2}{\zeta_0^2}\nonumber\\ &&\quad{}+ \Bigl(
\frac{R^2}{r_0^2}+1-\frac{2r_0^2}{R^2}\Bigr) \ln\frac{ R^2-r_0^2}{R^2}
\biggr]
\label{tffree}
\end{eqnarray}
where $\rho_0$ is the number density of the gas at the center of the
trap, and $\zeta_0$ is the corresponding healing length.

According to the implication that the gradient of the free energy
surface is related to the precession frequency as given in
Eq.~(\ref{findmotion}), one may calculate the precession frequency
within the approximations associated with Eq.~(\ref{tffree})
as~\cite{Lundh}
\begin{eqnarray}
  \boldvec{\omega} &=& \frac{\boldvec{\kappa}}{4\pi(R^2-r_0^2)}
  \Bigl(2\ln\frac{R}{\zeta_0} \nonumber\\
  &&\quad{}+\Bigl(\frac{R^4}{r_0^4}+2\Bigr) \ln\frac{
    R^2-r_0^2}{R^2}+\frac{R^2}{r_0^2}+2\Bigr)
\end{eqnarray}
Note that the precession direction for this case is always in the same
sense as the circulation, so that the qualitative behavior of the
motion of the vortex defect is similar to that of the uniform fluid in
a cylinder.

\begin{figure}
\begin{center}
\epsfysize=65mm \epsfbox{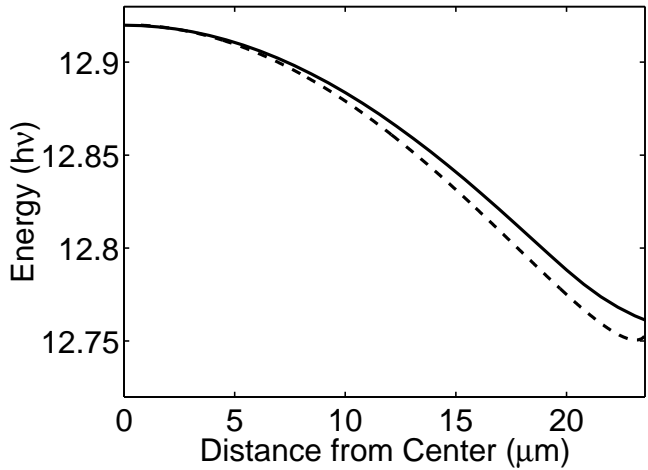}
\end{center}
\caption{Comparison of the exact numerical solution (solid line) for
  the free energy for the case of a Thomas-Fermi density distribution
  in the condensate with the analytic result derived in the
  hydrodynamic approximation given in Eq.~(\ref{tffree}) (dashed line)
  as the amount of displacement $r_0$ is varied. The energy falls off
  more gradually than for the previously considered case of a uniform
  fluid in a cylinder and there is no divergence of the analytic
  result at the boundary. The values of $N$, $a$, $m$, and $\nu$ are
  given in Fig.~\ref{HomoGPE}, with $\nu$ denoting in this case the
  frequency of the harmonic potential.}
\label{TFGPE}
\end{figure}

In Fig.~\ref{TFGPE} we compare the analytic expression for the free
energy given in Eq.~(\ref{tffree}) with the exact numerical solution
for the free energy. The agreement is remarkable and the role of edge
effects is small on the free energy and consequently on the precession
frequency. The sign of the free energy gradient indicates that in the
presence of dissipation, the vortex is not energetically stable and
will spiral outwards to the edge of the cloud where it will disappear.

\section{Two-Component Condensate}

When multiple-components are simultaneously present in a trap, the
interactions are characterized by a matrix of the interspecies and
intraspecies collisions. Depending on the elements of the scattering
matrix, various distinct kinds of behavior are possible for the
density distribution of the components.

Condensate experiments on $^{87}$Rb can typically trap two-components
simultaneously which tend to phase separate and are approximately
immiscible in a non-uniform confining potential. In addition, within
the mean-field approximation, the lowest energy solution has the
component with the smaller self-interaction scattering length in
regions of highest density. If this component is displaced it will
tend to float to the center of the trap, where in a harmonic potential
the density is greatest. A pictorial analogy is often made between
this physical mechanism and a buoyancy force in a fluid of varying
density.

In terms of the implications for the vortex dynamics, the inclusion of
this second component can be important even if the second component
contains no vortex lines itself. The extra degree of freedom
associated with the buoyancy behavior allows a more complex structure
for the free energy surface.

\begin{figure}
\begin{center}
\epsfysize=65mm \epsfbox{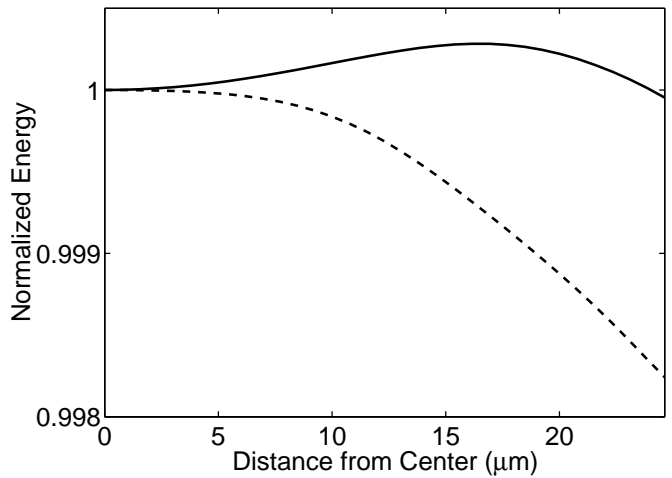}
\end{center}
\caption{Free energy for a two-component condensate as a ratio of
  the free energy when the defect is at the trap center. About 40\% of
  the atoms are in component~1 (the component containing no vortices)
  with the rest in component~2 (the component containing a single
  vortex line). The solid line is for $10^6$ total atoms and the
  dashed line is for $4\times 10^5$ total atoms.  With $10^6$ total
  atoms there is a minimum in the free energy at the trap center, and
  therefore an inner region of energetic stability. However, with
  $4\times 10^5$ total atoms, the mean-field effects are insufficient
  to generate the barrier.}
\label{BuoyGPE}
\end{figure}

In Fig.~\ref{BuoyGPE} we calculate the free energy for the case of
$10^6$ total $^{87}$Rb atoms in the condensate in an isotropic 7.8~Hz
trap with about 40\% in component~1 and 60\% in component~2. The
scattering parameter used is $g_{12}=4\pi\hbar^2(100a_0)/m$ where
$a_0$ is the Bohr radius. As mentioned previously the simulation is
actually performed in two-dimensions with a number of particles per
unit length set to give the same chemical potential as for the three
dimensional system.

The characteristic of a local minimum in the free energy at the center
of the trap has significant implications. The minimum implies a
critical radius $R_c$, which is the radius corresponding to the
maximum in the free energy curve. If the displacement of the vortex
line is less than $R_c$, the vortex will dissipate energy by spiraling
inwards to the center of the trap, where it will remain forever. There
is no energetic path by which the vortex can propagate to a region of
zero density and annihilate. Due to its non-trivial topological
structure, annihilation with a vortex of opposite circulation is
required to remove a vortex line and this can happen trivially only in
a zero-density region. Consequently, this two-component condensate
system studied here can support a persistent current which is
metastable, provided the vortex line is generated near the trap
center. In reality, inelastic processes which have not been included
in our discussion and which result from spin-exchange collisions will
limit the lifetime of the metastable state.

The complementary situation exists for a vortex line which is created
with a displacement larger than $R_c$. In the presence of dissipation,
such a vortex will spiral out of the system and annihilate at the
surface. If the vortex line is displaced from the trap center by
exactly $R_c$, the forces acting on the vortex defect balance
precisely, and the velocity of the singularity in the superfluid flow
is zero.

These phenomena require a sufficiently strong influence of the
mean-field interactions. If the total number of atoms is reduced from
$10^6$ to $4\times10^5$ with the same fraction in the non-rotating
component, there is no longer a barrier in the free energy, and the
critical radius is zero. In this case no energetically stable
persistent currents are possible.

\begin{figure}
\begin{center}
\epsfysize=65mm \epsfbox{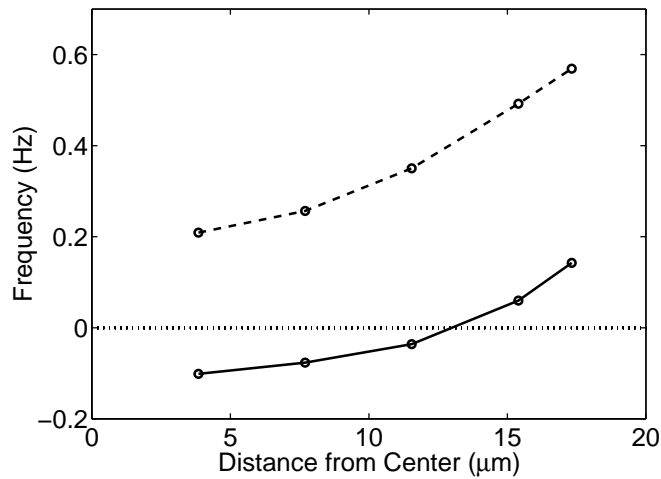}
\end{center}
\caption{The precession frequency as a function of displacement of the
defect for the two cases in Fig~\ref{BuoyGPE}.  In the region where
the gradient of the free energy points away from the center of the
trap, the precession is in the opposite sense (clockwise) to the sense
of the vortex fluid flow (counterclockwise).  Where the gradient of
the free energy points towards the center of the trap, the precession
is in the same sense as the vortex fluid flow.}
\label{Omega}
\end{figure}

The presence of a barrier in the free energy alters qualitatively the
behavior of the vortex dynamics. As illustrated in Fig.~\ref{Omega},
when the gradient of the free energy changes sign with respect to
displacement of the vortex core, so does the direction of precession
of the defect, as implied by Eq.~(\ref{findmotion}). Consequently, in
a two-component system, the non-rotating component can modify the rate
and direction of precession of the core purely through its influence
through the mean-field potential.

\begin{figure}
\begin{center}
\epsfysize=115mm \epsfbox{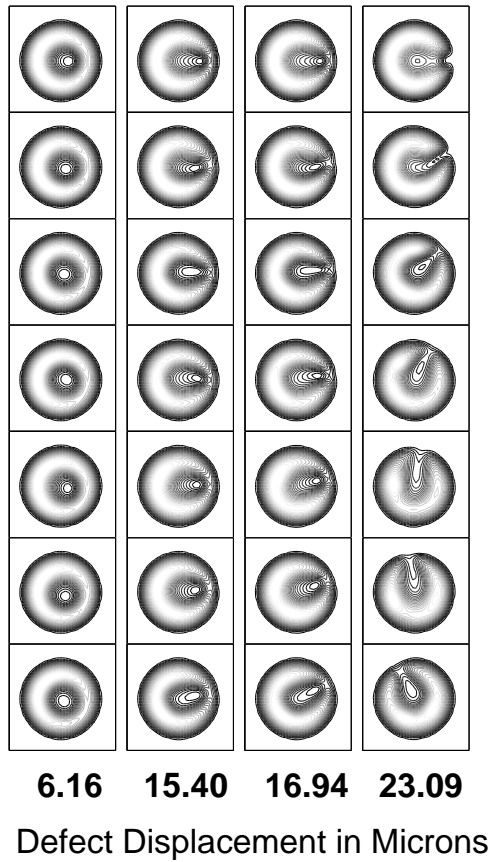}
\end{center}
\caption{Snapshots of the real-time precession of the defect for the
case of $10^6$ total atoms in a two-component condensate with
approximately 40\% in the non-rotating component. Each column
represents a time evolution from top to bottom with 100~ms between
each snapshot. The intrinsic superfluid circulation around the defect
has been chosen to be counterclockwise for all images. The initial
densities (top row) correspond to different displacements of the vortex
defect from the trap center. For the smallest displacement (left
column), the motion of the vortex defect is a clockwise precession
while the others are precessing counterclockwise. This is due to the
fact that only the first column corresponds to a displacement of the
vortex defect less than the critical radius given by the barrier in
the free energy for these parameters.}
\label{slides1}
\end{figure}

\begin{figure}
\begin{center}
\epsfysize=115mm \epsfbox{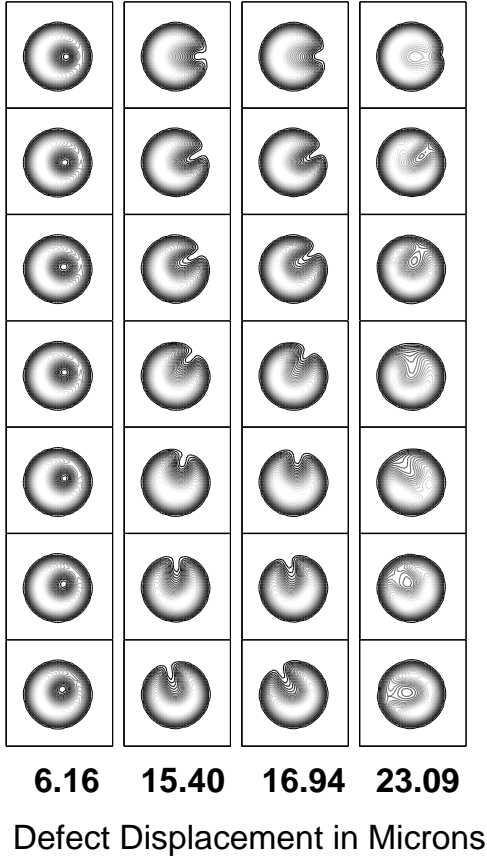}
\end{center}
\caption{Simulation of the real-time vortex dynamics as in
Fig.~(\ref{slides1}) with the same fraction of 40\% in the
non-rotating component, but illustrating the effect of reducing the
total number of atoms to $4\times 10^5$. The motion of the vortex
defect is in the same sense as the circulation of the vortex for all
displacements from the center. This results from the absence of the
free energy barrier due to the reduction in the mean-field buoyancy
effects.}
\label{slides2}
\end{figure}

In Fig.~\ref{slides1} we illustrate density snapshots showing
real-time images of the vortex dynamics for the case of $10^6$ total
atoms. The vortex defect in the first column, which has a displacement
less than the critical radius, is precessing clockwise, opposite to
the chosen direction of the superfluid circulation.  The others are
precessing counterclockwise, which is the same sense as the intrinsic
vortex fluid flow. For each of these columns, the rate of precession
is in accordance with the gradient of the of the free energy curve
calculated numerically for the same parameters. In Fig.~\ref{slides2}
similar snapshots are shown for the case for $4\times 10^5$ total
atoms.  Here, the defect is precessing counterclockwise at all
displacements, and the barrier in the free energy is absent.

\section{Conclusion}

We have found that for a single-component condensate the precession
direction will always be in the same sense as the vortex fluid flow
for both the uniform density in a cylinder and for the Thomas-Fermi
density profile. Such a vortex is not energetically stable, and in the
presence of dissipation, will spiral out to the edge of the cloud and
annihilate. Including the effects of kinetic energy in the core region
and edge effects at the surface of the cylinder in the free energy
gave minor modifications to the vortex dynamics from that predicted by
the hydrodynamic theories.

For a two-component condensate, numerical calculations demonstrated
the possibility for the precession direction to be opposite to the
sense of the vortex fluid flow. In this case, the vortex will
dissipate energy by spiraling to the center of the trap. Such a
situation occurs whenever there is a minimum in the free energy as a
function of displacement of the vortex defect. This possibility allows
a metastable persistent current which may remain indefinitely, even in
the presence of thermal fluctuations. This is due to the topological
nature of the vortex prohibiting annihilation and is a manifestation
of superfluidity. When there are insufficient atoms, the absence of an
energy barrier causes the vortex to spiral outward. This case is
similar to the case of a vortex in a single-component condensate, with
a modified precession frequency.

\section{Acknowledgements}

We would like to thank A. Fetter, E. Cornell, B. Anderson, and P.
Haljan for helpful discussions. This work was supported by the
National Science Foundation.

\end{document}